\documentclass[conference]{IEEEtran}
\IEEEoverridecommandlockouts
\usepackage{cite}
\usepackage{amsmath,amssymb,amsfonts}
\usepackage{algorithmic}
\usepackage{graphicx}
\usepackage{textcomp}
\usepackage{xcolor}
\usepackage{url}
\usepackage{tabularx}
\usepackage{booktabs}
\usepackage{lipsum}
\usepackage{multirow}
\usepackage{makecell}
\usepackage{amssymb}
\usepackage{booktabs}
\usepackage{float}
\usepackage{hyperref}
\usepackage{listings}
\usepackage{lstautogobble}

\definecolor{TUMBlue}{HTML}{0065BD}
\definecolor{TUMAccentGreen}{HTML}{A2AD00}

\def\BibTeX{{\rm B\kern-.05em{\sc i\kern-.025em b}\kern-.08em
    T\kern-.1667em\lower.7ex\hbox{E}\kern-.125emX}}
\begin{document}

\title{Hallucination in LLM-Based Code Generation: An Automotive Case Study}

\author{
\IEEEauthorblockN{Marc Pavel}
\IEEEauthorblockA{
\textit{Chair of Robotics, Artificial} \\ 
\textit{Intelligence and Real-Time Systems} \\
\textit{Technical University of Munich}\\
Munich, Germany \\
ge38raw@mytum.de}

\and
\IEEEauthorblockN{Nenad Petrovic}
\IEEEauthorblockA{
\textit{Chair of Robotics, Artificial} \\ 
\textit{Intelligence and Real-Time Systems} \\
\textit{Technical University of Munich}\\
Munich, Germany \\
nenad.petrovic@tum.de}
\and
\IEEEauthorblockN{Lukasz Mazur}
\IEEEauthorblockA{
\textit{Chair of Robotics, Artificial} \\ 
\textit{Intelligence and Real-Time Systems} \\
\textit{Technical University of Munich}\\
Munich, Germany \\
lukasz.mazur@tum.de}
\and
\IEEEauthorblockN{Vahid Zolfaghari}
\IEEEauthorblockA{
\textit{Chair of Robotics, Artificial} \\ 
\textit{Intelligence and Real-Time Systems} \\
\textit{Technical University of Munich}\\
Munich, Germany \\
v.zolfaghari@tum.de}
\and
\IEEEauthorblockN{Fengjunjie Pan}
\IEEEauthorblockA{
\textit{Chair of Robotics, Artificial} \\ 
\textit{Intelligence and Real-Time Systems} \\
\textit{Technical University of Munich}\\
Munich, Germany \\
f.pan@tum.de}
\and
\IEEEauthorblockN{Alois Knoll}
\IEEEauthorblockA{
\textit{Chair of Robotics, Artificial} \\ 
\textit{Intelligence and Real-Time Systems} \\
\textit{Technical University of Munich}\\
Munich, Germany \\
k@tum.de}
}

\maketitle

\begin{abstract}
Large Language Models (LLMs) have shown significant potential in automating code generation tasks offering new opportunities across software engineering domains. However, their practical application remains limited due to hallucinations - outputs that appear plausible but are factually incorrect, unverifiable or nonsensical. This paper investigates hallucination phenomena in the context of code generation with a specific focus on the automotive domain. A case study is presented that evaluates multiple code LLMs for three different prompting complexities ranging from a minimal one-liner prompt to a prompt with Covesa Vehicle Signal Specifications (VSS) as additional context and finally to a prompt with an additional code skeleton. The evaluation reveals a high frequency of syntax violations, invalid reference errors and API knowledge conflicts in state-of-the-art models GPT-4.1, Codex and GPT-4o. Among the evaluated models, only GPT-4.1 and GPT-4o were able to produce a correct solution when given the most context-rich prompt. Simpler prompting strategies failed to yield a working result, even after multiple refinement iterations. These findings highlight the need for effective mitigation techniques to ensure the safe and reliable use of LLM generated code, especially in safety-critical domains such as  automotive software systems.
\end{abstract}


\section{Introduction}
Over the last few years, Large language models (LLMs) have gained tremendous popularity in diverse domains such as software engineering, natural language processing and code generation. In particular, models such as GPT-3.5 \cite{3.5}, GPT-4 \cite{openai2024gpt4technicalreport}, CodeX \cite{codex}, CodeBERT \cite{codebert}, CodeGPT\cite{codegpt} and CodeLLaMA \cite{meta2023codellama} have become increasingly valuable due to their ability  to generate code snippets, refactor programs and provide programming assistance.

However, despite their impressive capabilities, LLMs have been observed to produce hallucinations - instances where the model generates information that contradicts its source, cannot be verified from it, or is presented as factual despite being nonsensical or incorrect \cite{surveyHallucination}\cite{hallucinationinllm}.

The general problem with hallucinations in LLMs lie in their tendency to generate output that lacks consistency and authenticity making the generated content unrealiable and unsuitable for practical applications \cite{llmhallucination}. This issue becomes even more critical in the context of code generation. Unlike natural language processing tasks where minor inaccuracies may be overlooked or easily corrected, errors in generated code can lead to bugs, security vulnerabilities, or system failures, potentially resulting in untrustworthiness, compromised data integrity, unauthorized access and increased development and maintenance costs \cite{lee2025hallucination}.

Hallucinations arise due to the statistical nature of LLMs, since they are trained to predict the most likely next token based on patterns in large datasets. However, they lack a true understanding of the meaning or the factual basis of the content they generate. As a result, when faced with ambiguity, sparsity of information, or requests beyond their knowledge, LLMs may fill in the gaps with plausible-sounding but incorrect or unverifiable content. Understanding how and why these failures occur is crucial for designing effective detection and mitigation strategies, especially in domains such as code generation. \cite{surveyHallucination}

The paper is organized as follows. The next section explores the nature and causes of LLM hallucinations in code generation. Moreover, the third section presents an automotive case study together with results, while appendix includes the list of vehicle signals used within the experiments. The last one summarizes the main findings of this work and potential future research directions.

\section{Hallucination in LLM-Based Code Generation}
While hallucinations in LLMs have been extensively studied, their occurence in code generation is an emerging challenge that has only recently begun to attract focus. Due to the complexity and diversity of programming tasks, existing classifications remain inconsistent with no universally accepted taxonomy for hallucinations in code generation. Moreover, current mitigation strategies are often task-specific and struggle to address the wide variety of hallucinations that can arise highlighting the need for more comprehensive solutions.

\subsection{Taxonomy}
To better understand and address hallucinations in code generation, recent studies have introduced systematic taxonomies. One such classification, proposed by Lee et al. \cite{lee2025hallucination}, highlights how these hallucinations can appear in various forms including bugs, syntactic errors, security vulnerabilities or even non-deterministic behavior:\newline

\begin{itemize}
    \item \textbf{Syntactic hallucinations:} These refer to hallucinations that cause compile-time failures and can be further categorized into \textbf{syntax violations}, such as missing brackets or incorrect indentation, and \textbf{incomplete code generation}, where the model produces unfinished code e.g. due to output length limitations.
    
    \item \textbf{Runtime execution hallucinations:} These refer to code that compiles correctly, but fails during execution. Runtime execution hallucinations inlcude \textbf{API knowledge conflicts}, which occur when the LLM misuses libraries or APIs e.g. by omitting necessary imports or using incorrect function signatures, and \textbf{invalid reference errors} where the generated code attempts to access or manipulate undefined elements, such as referencing undeclared variables or functions.
    
    \item \textbf{Functional correctness hallucinations:} These refer to code that compiles and executes without errors, but fails to meet the specified requirements. They can be further divided into \textbf{incorrect logical flow}, which includes missing corner cases, flawed conditional logic or incorrect arithmetic operations, and \textbf{requirement deviation}, where the generated code diverges from the explicit requirements and functionalities described in the prompt or problem description.
    
    \item \textbf{Code quality hallucinations:} These refer to code that introduces issues related to resource management, security or maintainability. This includes \textbf{resource mishandling} leading to excessive consumption or inefficient allocation of memory, \textbf{security vulnerabilities}, where generated code introduces weaknesses that make the system susceptible to attacks or unauthorized access and \textbf{code smells} referring to poorly structured or overly complex code that reducing maintainability.
\end{itemize}

\begin{table*}[ht]
\centering
\caption{Overview of the state-of-the-art code generation LLMs, their hallucination mitigation strategies, used benchmarks and ongoing challenges.}
\label{tab:overview-code}
\begin{tabularx}{\textwidth}{>{\raggedright\arraybackslash}p{3.5cm} >{\raggedright\arraybackslash}X >{\raggedright\arraybackslash}p{3cm} >{\raggedright\arraybackslash}p{3.2cm}}
\toprule
\textbf{Model} & \textbf{Hallucination Mitigation Techniques} & \textbf{Benchmarks} & \textbf{Challenges} \\
\midrule

GPT-4.1 (2025) \cite{openai2025gpt41} 
& Instruction tuning, long context up to 1 million tokens
& SWE-Bench, Graphwalks, MultiChallenge, etc.
& Limited retrieval grounding \\

Codex (2021) \cite{codex}
& Fine-tuned on GPT-3 with code from GitHub
& HumanEval
& Difficulty with docstrings describing long chains of operations and with binding operations to variables \\

Claude 3 Opus (2024) \cite{anthropic2024claude3}
& Constitutional AI, RLHF, red-teaming 
& HumanEval, MBPP, APPS
& No web search, factual errors and bias \\

Code LLaMA 70B (2024) \cite{meta2023codellama}
& Infilling, long input contexts, instruction fine-tuning
& HumanEval, MBPP, MultiPL-E
& Limited retrieval grounding \\

DeepSeek-Coder V2 (2024) \cite{deepseek2024coderv2}
& long input contexts, RL, instruction fine-tuning
& HumanEval, MBPP+, SWE-Bench, etc.
& Gap in instruction-following capabilities \\

StarCoder2 15B (2024) \cite{starcoder22024}
& No native mitigation
& HumanEval, MBPP, EvalPlus, etc.
& Representation bias, societal bias \\

Phi-3 Mini (2024) \cite{microsoft2024phi3}
& Post-processing including supervised fine-tuning and direct preference optimization
& HumanEval, MBPP, DS-1000, etc.
& Limited capacity to store factual knowledge due to model size \\

WizardCoder 34B (2023) \cite{xu2023wizardcoder}
& Instruction tuning via Evol-Instruct
& HumanEval, MBPP 
& Tendency to lose semantic and logical consistency during code generation \cite{tian2025codehaluinvestigatingcodehallucinations}\\

\bottomrule
\end{tabularx}
\end{table*}

\subsection{Causes}
To understand these hallucinations more thoroughly, recent research has examined their root causes, commonly classifying them into three main categories \cite{lee2025hallucination}:

\begin{itemize}
    \item \textbf{Training data issues:} These refer to limitations in the dataset including the lack of quantity of training data and the low quality of training datasets.

    \item \textbf{Trained model issues:} These refer to problems within the model itself such as the lack of proper evaluation methods, limited ability to reason or understand complex logic, non-deterministic output due to the temperature settings or constraints regarding maximum input token length.
    
    \item \textbf{Prompt issues:} These refer to unclear or poorly constructed user input such as ambiguous prompts or irrelevant context in prompts.
\end{itemize}

\subsection{Mitigation}
Mitigating hallucinations in code generation is a complex challenge, where no one-size-fits-all solution exists. Consequently, a variety of strategies have been developed, each targeting specific types of hallucinations. 

\subsubsection{De-Hallucinator}
The De-Hallucinator, proposed by Eghbali et al. \cite{eghbali2024dehallucinator}, addresses two key challenges in code generation: identifying which parts of the project code are essential to include in the prompt and mitigating function misuse due to the LLM's lack of knowledge of project-specific APIs. Firstly, their approch pre-analyses and indexes the source code of the project. When a user submits an initial prompt, relevant APIs are then being retrieved using RAG. To further reduce hallucinations, they also proposed to apply this method iteratively. The final context enhanced prompt is then used as input for the LLM-based code generation.

\subsubsection{ClarifyGPT}
ClarifyGPT, proposed by Mu et al. \cite{ClarifyGPT}, aims to reduce hallucinations caused by ambiguous prompts. The approach firstly generates diverse test inputs, followed by a code consistency check that compares the outputs of multiple generated code solutions on the same input to identify unclear requirements. If ambiguity is detected, the framework then generates reasoning based questions, which are being prompted to the user. Finally, the prompt is being refined based on the users response and queried to the LLM for the enhanced code generation. 

\subsubsection{Refining ChatGPT-Generated Code}
The framework refining ChatGPT-generated Code by Liu et al. \cite{liu2023refiningchatgptgeneratedcodecharacterizing} aims to mitigate all code quality issues in the generated code. The core idea is to prompt the LLM to refine the generated code once quality issues are detected. In the paper, they proposed two different repairing prompt types: (1) simple feedback telling the LLM that the generated code has quality issues without giving detailed error information; (2) feedback with static analysis and runtime errors providing detailed feedback to guide code revision in the correct direction. If these refinements do not resolve the code quality issues, the framework applies iterative repairing to enhance the code through multiple feedback cycles.

\subsection{Overview}
To complement the discussion on taxonomy, causes and mitigation strategies for hallucinations in code generation, Table \ref{tab:overview-code} provides an overview of current state-of-the-art models used for code generation. The table outlines their respective hallucination mititgation techniques, used benchmarks and remaining challenges.

\section{Automotive Case Study}
To better understand the extent of hallucinations, especially in unseen domains, an automotive case study was conducted. The goal was to assess the reliability by observing the types and frequency of hallucinations and to evaluate the effectiveness of a mitigation strategy. 

\subsection{Experimental Setting}

\subsubsection{Task Description}
The task was to generate code, which turns off the windshield wipers, when the hood of the vehicle is open. To implement this, we made use of the COVESA Vehicle Signal Specification\footnote{\url{https://covesa.github.io/vehicle\_signal\_specification/introduction/overview/}} (VSS), which provides a standardized set of signals for various vehicle components. The relevant VSS signals for this task are:

\begin{itemize}
    \item \textbf{Vehicle.Body.Hood.IsOpen:} Used to check whether the hood is open.
    \item \textbf{Vehicle.Body.Windshield.Front.Wiping.Mode:} Used to turn off the front windshield wipers.
\end{itemize}

\subsubsection{Models and Mitigation Strategy}
The task was evaluated across seven offline code generation LLMs including StarCoder (1B), TinyLLaMA, CodeLLaMA (7B, 7B-python, 7B-instruct), Phi-2 and Deepseek-Coder as well as three online LLMs being GPT-4o, GPT-4.1 and Codex. 
To mitigate hallucinations, we applied the iterative repairing technique with feedback with static analysis and runtime errors proposed in Refining ChatGPT-generated code \cite{liu2023refiningchatgptgeneratedcodecharacterizing}.

\subsubsection{Prompting Strategies}
Each model was tested on three different prompting complexities:

\begin{enumerate}
    \item \textbf{Baseline:} a minimal one-liner prompt specifying the task.
    \item \textbf{Signal-augmented prompt:} baseline prompt plus a list of 20 potentially relevant VSS signals
    \item \textbf{Template-augmented prompt:} signal-augmented prompt plus a code skeleton with TODOs 
\end{enumerate}

\subsubsection{Evaluation Metric}
To evaluate the correctness of the generated code, we define a heuristic scoring metric that starts at a maximum of 1.0 and penalizes occuring hallucinations in the following way:

\begin{itemize}
    \item \textbf{-0.2:} compilation or runtime errors (missing async routines, missing imports or referencing undefined variables)
    \item \textbf{-0.1:} Use of wrong VSS signals or improper manipulation of VSS signals.
\end{itemize}

\subsection{Baseline prompt}
\textbf{Prompt:}
Generate a full python code example using SDV python SDK and Covesa VSS signals to turn off the wipers as soon as the hood is open.\newline

\begin{table}[H]
\centering
\caption{Baseline prompt using GPT-4o}
\label{B4o}
\begin{tabular}{p{0.3cm}p{0.3cm}p{1.5cm}p{3.5cm}p{0.5cm}}
\toprule
\textbf{It} & \textbf{Suc.}  & \textbf{Taxonomy} & \textbf{Description} & \textbf{Score} \\
\midrule
0 & $\times$ & SV, IRE, API KConflict & DataPointChange from sdv.model & 0.1 \\
1 & $\times$ & SV, IRE, API KConflict & sdv.databroker & 0.1 \\
2 & $\times$ & IRE, API KConflict & Vehicle.\_\_init\_\_() missing argument & 0.5 \\
3 & $\times$ & IRE, API KConflict & velocitas.\_sdk. model.datapoint & 0.5 \\
4 & $\times$ & IRE, API KConflict & Wiping object has no attribute SystemState & 0.5 \\
\bottomrule
\end{tabular}
\end{table}

Table \ref{B4o}, \ref{B41} and \ref{BC} show the progression of closeness scores over five repair iterations for GPT-4o, GPT4.1 and Codex respectively. Initial scores (0.1, 0.2, 0.2) increased only modestly (to 0.5, 0.4, 0.4) with more iterations indicating that while the mitigation strategy improved the generated code, it was insufficient to solve the task. In some iterations, the score got even worse as solving one error caused multiple new ones to arise. The recorded types of hallucinations are syntax violations, invalid reference errors and API knowledge conflicts. Notably, all three models hallucinated VSS signals that do not exist indicating a reliability issue in generating code for unfamiliar domains.

\begin{table}[H]
\centering
\caption{Baseline prompt using GPT-4.1}
\label{B41}
\begin{tabular}{p{0.3cm}p{0.3cm}p{1.5cm}p{3.5cm}p{0.5cm}}
\toprule
\textbf{It} & \textbf{Suc.}  & \textbf{Taxonomy} & \textbf{Description} & \textbf{Score} \\
\midrule
0 & $\times$ & SV, IRE, API KConflict & SDVApp from sdv & 0.2 \\
1 & $\times$ & SV, IRE, API KConflict & run\_app from sdv.vehicle\_app & 0.0 \\
2 & $\times$ & IRE, API KConflict & WiperControlApp object has no attribute vehicle & 0.2 \\
3 & $\times$ & IRE, API KConflict & WiperControlAPp object has no attribtue datapoints & 0.2 \\
4 & $\times$ & IRE, API KConflict & Vehicle.\_\_init\_\_() missing argument & 0.4 \\
\bottomrule
\end{tabular}
\end{table}

\begin{table}[H]
\centering
\caption{Baseline prompt using Codex}
\label{BC}
\begin{tabular}{p{0.3cm}p{0.3cm}p{1.5cm}p{3.5cm}p{0.5cm}}
\toprule
\textbf{It} & \textbf{Suc.}  & \textbf{Taxonomy} & \textbf{Description} & \textbf{Score} \\
\midrule
0 & $\times$ & SV, IRE, API KConflict & subscribe\_signal form sdv.vehicle\_app & 0.2 \\
1 & $\times$ & SV, IRE, API KConflict & subscribe\_to\_signal from sdv.vehicle\_app & 0.0 \\
2 & $\times$ & SV, IRE, API KConflict & Vehicle from sdv.model & 0.2 \\
3 & $\times$ & SV, IRE, API KConflict & No async routine & 0.2 \\
4 & $\times$ & IRE, API KConflict & vehicle has no attribute Body & 0.4 \\
\bottomrule
\end{tabular}
\end{table}

\subsection{Signal-augmented prompt}
\textbf{Prompt:}
Generate a full python code example using SDV python SDK and some of the provided Covesa VSS signals to turn off the wipers as soon as the hood is open (the list of provided VSS signals is available in the Appendix \ref{section:vss}).

Tables \ref{S4o}, \ref{S41} and \ref{SC} display the results for the signal-augmented prompt. Compared to the baseline, all models start with higher initial scores (0.4, 0.3, 0.5) and end with moderately improved results (to 0.4, 0.7, 0.7) while still producing the same hallucination types as before. Moreover, hallucinations of non-existent VSS signals are no longer observed. However, all models struggle to differentiate between similar signals, which means they use Vehicle.Body.Windshield.Front.Wiping.System.Model instead of the correct one. This suggests that while providing signal lists help narrow down the possible outputs, distinguishing between similar APIs still remains a challenge.

\begin{table}[H]
\centering
\caption{Signal-augmented prompt using GPT-4o}
\label{S4o}
\begin{tabular}{p{0.3cm}p{0.3cm}p{1.5cm}p{3.5cm}p{0.5cm}}
\toprule
\textbf{It} & \textbf{Suc.}  & \textbf{Taxonomy} & \textbf{Description} & \textbf{Score} \\
\midrule 
0 & $\times$ & IRE, API KConflict & sdv.sdk & 0.4 \\ 
1 & $\times$ & SV, IRE, API KConflict & vehicle\_app & 0.0 \\ 
2 & $\times$ & IRE, API KConflict & AppMetadata from sdv.vehicle\_app & 0.2 \\ 
3 & $\times$ & IRE, API KConflict & vehicle from sdv.model & 0.4 \\ 
4 & $\times$ & IRE, API KConflict & WipperControlApp object has no attribute logger & 0.4 \\
\bottomrule
\end{tabular}
\end{table}

\begin{table}[H]
\centering
\caption{Signal-augmented prompt using GPT-4.1}
\label{S41}
\begin{tabular}{p{0.3cm}p{0.3cm}p{1.5cm}p{3.5cm}p{0.5cm}}
\toprule
\textbf{It} & \textbf{Suc.}  & \textbf{Taxonomy} & \textbf{Description} & \textbf{Score} \\
\midrule
0 & $\times$ & SV, IRE, API KConflict & subscribe\_signal from sdv.vehicle\_app & 0.3 \\
1 & $\times$ & SV, IRE, API KConflict & Vehicle from sdv.model & 0.5 \\
2 & $\times$ & SV, IRE, API KConflict & No async routine & 0.5 \\
3 & $\times$ & IRE, API KConflict & WiperSafetApp object has no attribute vehicle & 0.7 \\
4 & $\times$ & IRE, API KConflict & WiperControlApp object has no attribute logger & 0.7 \\
\bottomrule
\end{tabular}
\end{table}

\begin{table}[H]
\centering
\caption{Signal-augmented prompt using Codex}
\label{SC}
\begin{tabular}{p{0.3cm}p{0.3cm}p{1.5cm}p{3.5cm}p{0.5cm}}
\toprule
\textbf{It} & \textbf{Suc.}  & \textbf{Taxonomy} & \textbf{Description} & \textbf{Score} \\
\midrule
0 & $\times$ & IRE, API KConflict & SDVClient from sdv & 0.5 \\
1 & $\times$ & SV, IRE, API KConflict & Vehicle from sdv.model & 0.5 \\
2 & $\times$ & SV, API KConflict & No async routine & 0.5 \\
3 & $\times$ & API KConflict & Wipers are still on, when hood is open & 0.7 \\
4 & $\times$ & IRE, API KConflict & HoodWiperController object has no attribute vehicle & 0.7 \\
\bottomrule
\end{tabular}
\end{table}

\subsection{Template-augmented prompt}
\textbf{Prompt:}
Fill in the code skeleton (see Listing \ref{fig:codeTemplate}) using SDV python SDK and Covesa VSS signals to turn off the wipers as soon as the hood is open. (the list of provided VSS signals is available in the Appendix \ref{section:vss})

\begin{lstlisting}[language=Python,caption=Code Template used in the template-augmented prompt, captionpos=b, label=fig:codeTemplate]
import time
import asyncio
import signal
from sdv.vdb.reply import DataPointReply
from sdv.vehicle_app import VehicleApp
from vehicle import Vehicle, vehicle

class TestApp(VehicleApp):
    def __init__(self, vehicle_client: Vehicle):
        # TODO code

    async def on_start(self):
        # TODO code

async def main():
    vehicle_app = TestApp(vehicle)
    await vehicle_app.run()

LOOP = asyncio.get_event_loop()
LOOP.add_signal_handler(signal.SIGTERM, LOOP.stop)
LOOP.run_until_complete(main())
LOOP.close()
\end{lstlisting}

\begin{table}[H]
\centering
\caption{Template-augmented prompt using GPT-4o}
\label{T4o}
\begin{tabular}{p{0.3cm}p{0.3cm}p{1.5cm}p{3.5cm}p{0.5cm}}
\toprule
\textbf{It} & \textbf{Suc.}  & \textbf{Taxonomy} & \textbf{Description} & \textbf{Score} \\
\midrule 
0 & $\times$ & API KConflict & set target value for non-actuator is not allowed & 0.8 \\ 
1 & $\times$ & API KConflict & wrong signal: Vehicle.Body. Windshield.Front. Wiping.System.Mode & 0.9 \\ 
2 & \checkmark & - & - & 1.0 \\ 

\bottomrule
\end{tabular}
\end{table}

Tables \ref{T4o}, \ref{T41} and \ref{TC} display the results for the template-augmented prompt. Compared to the baseline and the signal-augmented prompt, all models achieve significantly higher initial scores (0.8, 0.9, 0.9) 
with the only remaining issues being API knowledge conflicts such as usage of incorrect VSS signals or improper manipulation of VSS signals. For GPT-4o and GPT-4.1, the mitigation strategy was effective: GPT-4o succeeded after two iterations, while GPT-4.1 produces a correct solution after only 1 iteration. This demonstrates that when the model is guided with both a structured starting point (code template) and a constrained set of signals, the Refining ChatGPT-generated code \cite{liu2023refiningchatgptgeneratedcodecharacterizing} technique can lead to a complete solution. However, Codex maintained a score of 0.9 across all iterations and was unable to produce a correct solution. After the second iteration, it declared the task unsolvable given the provided VSS signal list. This emphasizes the need for caution when relying on generated code in unfamiliar domains, because even with structured prompts and iterative feedback, correcting errors is not always straightforward and successful repair is not guaranteed. The seven Offline code generation models were unable to generate any sensible code, highlighting their unsuitability for domain-specific tasks they sere not trained on even with context-enriched prompts.

\begin{table}[H]
\centering
\caption{Template-augmented prompt using GPT-4.1}
\label{T41}
\begin{tabular}{p{0.3cm}p{0.3cm}p{1.5cm}p{3.5cm}p{0.5cm}}
\toprule
\textbf{It} & \textbf{Suc.}  & \textbf{Taxonomy} & \textbf{Description} & \textbf{Score} \\
\midrule
0 & $\times$ & API KConflict & setting int for string variable & 0.9 \\
1 & \checkmark & - & - & 1.0 \\
\bottomrule
\end{tabular}
\end{table}

\begin{table}[H]
\centering
\caption{Template-augmented prompt using Codex}
\label{TC}
\begin{tabular}{p{0.3cm}p{0.3cm}p{1.5cm}p{3.5cm}p{0.5cm}}
\toprule
\textbf{It} & \textbf{Suc.}  & \textbf{Taxonomy} & \textbf{Description} & \textbf{Score} \\
\midrule
0 & $\times$ & API KConflict & wrong signal: Vehicle.Body. Windshield.Front. Wiping.System.Mode & 0.9 \\
1 & $\times$ & API KConflict & Wipers are still on, when hood is open & 0.9 \\
2 & $\times$ & API KConflict & Wipers are still on, when hood is open & 0.9 \\
\bottomrule
\end{tabular}
\end{table}

\section{Conclusion}
In this study, we conducted an automotive case study applying Covesa VSS signals to a real-world task: generate pyhton code to turn off the windshield wipers when the hood is open. We explored multiple prompting strategies, namely baseline, signal-augmented and template-augmented, and applied the Refining ChatGPT-generated Code \cite{liu2023refiningchatgptgeneratedcodecharacterizing} approach to iteratively provide feedback to improve code generation quality.
Regarding reliability, the baseline and signal-augmented prompts frequently led to syntax violations, invalid reference errors and API knowledge conflicts. While syntax violations were often resolved through iterative feedback, the other hallucinations persisted. On the other hand, the template-augmented prompt produced only API knowledge conflicts. The most critical hallucinations were hallucinations of non-existent VSS signals for the baseline prompt and choosing the correct signal amongst similar ones for the other two prompting techniques.

Regarding the effectiveness of the mitigation strategy, the scores improved consistently across all prompting strategies demonstrating general effectiveness. However, a working sample solution was only achieved with the template-augmented prompt for GPT-4o and GPT-4.1 indicating limitations in how far the strategy can go depending on the prompt structure and model capabilities.

These results suggest that different types of hallucinations in code generation require tailored mitigation strategies. Since this is a new area, not all types of hallucinations were explored in detail hinting at the need of more research to close the remaining gap. As future work, this work could be extended to more online LLMs as well as different mitigation strategies to better understand and improve reliability and effectiveness in code generation. Finally, we would also consider integration with our previous work on synergy of formally grounded model-driven engineering and LLM based methods for autonomous driving capabilities code\cite{petrovic2024llmdriventesting} and test scenario generation \cite{lebioda2025requirements} starting from textual requirements.

\appendices
\section{Covesa VSS Signals}\label{section:vss}
The following COVESA VSS signals were provided as input in prompting technique 2 and 3. They comprise all signals related to the hood of the vehicle and the front windshield wipers.

\begin{itemize}
    \item Vehicle.Body.Hood
    \item Vehicle.Body.Hood.IsOpen
    \item Vehicle.Body.Hood.Position
    \item Vehicle.Body.Hood.Switch
    \item Vehicle.Body.Windshield.Front.Wiping
    \item Vehicle.Body.Windshield.Front.Wiping.Intensity
    \item Vehicle.Body.Windshield.Front.Wiping.IsWipersWorn
    \item Vehicle.Body.Windshield.Front.Wiping.Mode
    \item Vehicle.Body.Windshield.Front.Wiping.System
    \item Vehicle.Body.Windshield.Front.Wiping.System.ActualPosition
    \item Vehicle.Body.Windshield.Front.Wiping.System.DriveCurrent
    \item Vehicle.Body.Windshield.Front.Wiping.System.Frequency
    \item Vehicle.Body.Windshield.Front.Wiping.System.IsBlocked
    \item Vehicle.Body.Windshield.Front.Wiping.System.IsEndingWipeCycle
    \item Vehicle.Body.Windshield.Front.Wiping.System.IsOverheated
    \item Vehicle.Body.Windshield.Front.Wiping.System.IsPositionReached
    \item Vehicle.Body.Windshield.Front.Wiping.System.IsWiperError
    \item Vehicle.Body.Windshield.Front.Wiping.System.IsWiping
    \item Vehicle.Body.Windshield.Front.Wiping.System.Mode
    \item Vehicle.Body.Windshield.Front.Wiping.System.TargetPosition
    \item Vehicle.Body.Windshield.Front.Wiping.WiperWear
\end{itemize}

\end{document}